\documentclass[showpacs,aps,amsmath,amssymb,twocolumn]{revtex4}
\usepackage[latin1]{inputenc}
\usepackage{dcolumn}
\usepackage{bm}
\usepackage{verbatim} 
\usepackage[]{graphicx}			
\usepackage{color}			
\usepackage{subfig}			
\graphicspath{{./figuras/}}		
\usepackage{caption}	
\newcommand{\be}{\begin{equation}}
\newcommand{\ee}{\end{equation}}
\newcommand{\ben}{\begin{eqnarray}}
\newcommand{\een}{\end{eqnarray}}
\newcommand{\bes}{\begin{subequations}}
\newcommand{\ees}{\end{subequations}}
\newcommand{\bb}{\bibitem}

\newcommand{\bfi}{\begin{figure}}
\newcommand{\efi}{\end{figure}}
\newcommand{\bc}{\begin{center}}
\newcommand{\ec}{\end{center}}
\newcommand{\sech}{\mbox{sech}}
\begin{document}
\title{Twinlike models for kinks and compactons in flat and warped spacetime}

\author{D. Bazeia$^{1,2,3}$, A.S. Lob\~ao Jr.$^{2}$, and R. Menezes$^{3,4}$}

\affiliation{$^1$Instituto de F\'\i sica, Universidade de S\~ao Paulo, 05314-970, S\~ao Paulo, SP, Brazil\\$^2$Departamento de Física, Universidade Federal da Para\'\i ba, 58051-970 Jo\~ao Pessoa, PB, Brazil\\ $^3$Departamento de F\'\i sica, Universidade Federal de Campina Grande, 58109-970 Campina Grande, PB, Brazil and \\ $^4$Departamento de Ci\^encias Exatas, Universidade Federal da Para\'\i ba, 58297-000 Rio Tinto, PB, Brazil.}

\begin{abstract}
This work deals with the presence of twinlike models in scalar field theories. We show how to build distinct scalar field theories having the same extended solution, with the same energy density and the very same linear stability. Here, however, we start from a given but generalized scalar field theory, and we construct the corresponding twin model, which also engenders generalized dynamics. We investigate how the twinlike models arise in flat and in curved spacetime. In the curved spacetime, we consider a braneword model with the warp factor controlling the spacetime geometry with a single extra dimension of infinite extent. In particular, we study linear stability in the flat and curved spacetime, and in the case of curved spacetime, in both the gravity and the scalar field sectors, for the two braneworld models.
\end{abstract}


\maketitle
%%%%%%%%%%%%%%%%%%%%%%%%%%%%%%%%%%%%%%%%%%%%%%%%%%%%%%%%%%%%%%%%%%%%%

%%%%%%%%%%%%%%%%%%%%%%%%%%
\section{introduction}

Topological structures are of great interest in high energy physics \cite{b1,b2} and in other areas of nonlinear science \cite{b3,b4,b5}. In high energy physics, they are kinks, vortices, monopoles and other field configurations. They are finite energy field configurations which solve the equations of motion of the respective models in the corresponding spacetime dimensions.  Kinks are the simplest structures, and they are usually constructed under the presence of real scalar fields in $(1,1)$ spacetime dimensions. 

In this work we focus on the presence of kinks in model described by real scalar fields, driving attention to the recent issue concerning the investigation of twinlike models. This issue was firstly considered in \cite{altw}, and then studied in a diversity of contexts in the works \cite{t1,t2,t3,t31,t4}.
In the papers on twinlike models \cite{altw,t1,t2,t3,t31,t4}, the key issue was to construct and investigate twinlike models, starting from a standard model and then introducing the twin model, which usually appears describing generalized or non-standard dynamics. The important point is that it is sometimes possible to find a standard and another model, with generalized dynamics, both having the same defect solution, with the very same energy density. The two models are then twinlike models. 
However, in Ref.~\cite{t3} some of us have shown that it is possible to have twinlike models with the very same stability behavior. We call this the strong condiction: that is, there are twin models, if they have the same solution, with the very same energy density; but there are models that are twins in the strong sense, if they also have the very same behavior, concerning linear stability.

To enlarge the scope of the work, we also investigate models of the Randall-Sundrum type \cite{RS}, in the presence of scalar fields, as suggested in Ref.~\cite{GW}. The issue was investigated in several works \cite{first11,first22}, and here we consider the scalar field with generalized dynamics, leading us to a mathematical framework which is much more complicated then it appears in the case of standard dynamics. In spite of this, we introduce a complete investigation of linear stability, both in the flat and curved spacetime.

The main issue here is to open another route to deal with twinlike models. Indeed, we consider the construction of twinlike models, but now we start from a generalized model, instead of using a standard field theory. The issue is of current interest, mainly because models with generalized dynamics have been used to respond for the presence of dark energy and dark matter, and to test possible modifications of general relativity. However, this is not a simple question, because of the intricacy of the the models to be investigated. To ease the investigation, we follow Refs.~\cite{first1,first2,fb}. In particular, we introduce a new function $W=W(\phi)$, from which we obtain simple first-order equations that very much help us to study and solve the equations of motion. The presence of $W(\phi)$ allows for  supersymmetric extensions, as previouslly investigated in \cite{Bazeia:2009db,Adam:2011gc}. 

We organize the current work as follows: in Sec.~\ref{gen} we introduce the procedure, starting from a generalized model. In Sec.~\ref{other} we investigate other generalized models, to show that the procedure is general, and works not only for the model introduced in the previous section. In Sec.~\ref{brane} we extend the procedure to the braneworld scenario, in the case of warped geometry with a single extra dimension of infinite extent. Finally, in Sec.~\ref{end} we end the work with some comments and conclusions.

%%%%%%%%%%%%%%%%%%%%
\section{The new procedure}\label{gen}

Let us start following the lines of \cite{t1}. We consider the model described by a single real scalar field $\phi$, with the non standard Lagrange density
\begin{equation}\label{eq1}
{\cal L}=\frac{2^{n-1}}{n}X|X|^{n-1}-U(\phi)
\end{equation}
where 
$$
X\equiv \frac{1}{2}\partial_\mu\phi\partial^\mu\phi
$$ 
and $U(\phi)$ is the potential that identifies the theory. In this work we deal with bidimensional spacetime, with metric $ds^2=dt^2-dx_1^2$, using $x_0=t$ and $x^1=-x_1=x$. Here we take $\hbar=c=1$ and we assume that the field and coordinates are all dimensionless.

The equation of motion for this theory is
\begin{equation}\label{eq2}
2^{n-1}\partial_\mu(|X|^{n-1}\partial^\mu\phi)+U_\phi=0
\end{equation}
where $U_\phi=dU/d\phi$ and the energy-momentum tensor is given by
\begin{equation}
T_{\mu\nu}=-g_{\mu\nu}{\cal L}+{\cal L}_X\partial_\mu \phi\partial_\nu\phi,
\end{equation}
where ${\cal L}_X=\partial {\cal L}/\partial X$.
Here we are interested in static solutions, $\phi=\phi(x)$, so we have that
\begin{eqnarray}
T_{00}&=&\rho=\frac{1}{2n}\phi^{\prime 2n}+U,\\
T_{11}&=&\frac{(2n-1)}{2n}\phi^{\prime 2n}-U(\phi).
\end{eqnarray}
Moreover, the equation of motion $(\ref{eq2})$ becomes
\begin{equation}\label{eq3}
\left(\phi^{\prime 2n-1}\right)^{\prime}=U_\phi|_{\phi=\phi_s(x)}.
\end{equation}
where $|_{\phi=\phi_s}$ indicates that we have to consider the field static, that is, $\phi=\phi(x)$. This fact will be denoted from now on by $|_s$. This equation can be integrated once and we obtain
\begin{equation}\label{eq4}
\frac{(2n-1)}{2n}\phi^{\prime 2n}-U(\phi)|_s=C,
\end{equation}
where $C$ is a constant that can be identified with the stress tensor $T_{11}$. Stability  of  the  static  solution imposes that $C=0$, and this makes the energy density $T_{00}$ to get to the form
\bes
\begin{equation}\label{eq5}
\rho(x)=\phi^{\prime 2n},
\end{equation} 
or
\begin{equation}\label{eqp5}
\rho(x)=\frac{2n}{2n-1}U(\phi(x)).
\end{equation}  
\ees
Let us now follow the procedure introduced in \cite{fb} to write the equation \eqref{eq5} in another form, much more convenient to study braneword models. The key step is to introduce a new function, $W=W(\phi)$, such that
\begin{equation}\label{eq6}
{\cal L}_X \phi^{\prime}=W_\phi.
\end{equation}
This fact leads us to
\begin{equation}\label{eq7}
\phi^{\prime}=W_\phi^{\frac{1}{2n-1}}
\end{equation}
and so we can write
\begin{equation}\label{eq8}
U(\phi)=\frac{2n-1}{2n}W_\phi^{\frac{2n}{2n-1}}.
\end{equation}
Therefore, the energy density has the form
\begin{equation}\label{eq9}
\rho(x)=W_\phi^{\frac{2n}{2n-1}}
\end{equation}
We illustrate this procedure introducing two choices for the function $W(\phi)$. First, we choose the following $n-$dependent function 
\be\label{spo1}
W(\phi)=\phi\,\mbox{}_{2}F_1\left(\frac12,-2n+1;\frac32;\phi^2\right)
\ee
this is a polynomial function of degree $4n-1$. The presence of the hipergeometric function $\mbox{}_{2}F_1$ introduces a new and general form to write $W(\phi)$; for example, we can write
\bes\label{W1eq}
\ben
W(\phi)&=&\phi - \frac13 \phi^3 \\
W(\phi)&=&\phi-\phi^{ 3}+\frac{3}{5}\phi^5-\dfrac{1}{7}\phi^7 \label{sp1}
\een
\ees
for $n=1$ and $n=2$, respectively. 

In Ref. \cite{first2}, it was shown that the two choices given by Eqs. \eqref{W1eq} lead us with kinklike solution. Here we investigated the behavior of the model characterized by the general parameter  $n$.

Using the Eq.~\eqref{eq8}, we obtain the potential
\be
U(\phi)=\frac{2n-1}{2n}\left(1-\phi^2\right)^{2n}\label{pott1}
\ee
For the first potential the equation \eqref{eq7} can be write as $\phi^{\prime}=1-\phi^2$, whose the solution is 
\be
\phi(x)=\tanh(x).
\ee 
The energy density is given by 
\be
\rho(x)=\sech^{4n}(x)
\ee
The solution despite not having dependency of the parameter, the energy density depends. The behavior asymptotic for different values for n is 
\be\label{asympt}
\rho(x) = 2^n e^{-nx}+\ldots
\ee
Is is clear that the solutions and energy densities asymptotic
behavior slow down with an increasing $n$. The energy is 
\be
E=\frac{\sqrt{\pi}\,\Gamma{\left(2n\right)}}{\Gamma\left(2n+\frac12\right)}
\ee 
For $n=1$ and $n=2$, the energy is $E=4/3$ and $E=32/35$, respectively.

Now we introduce the second function, given by
\be
W(\phi)=\phi\,\mbox{}_{2}F_1\left(\frac12,-n+\frac12;\frac32;\phi^2\right) \label{numvi}
\ee
For example, we can write this function
\bes
\begin{eqnarray}
W(\phi)\!&=&\!\frac{\phi}{2}\sqrt{|1-\phi^2|}+\frac12 \arcsin(\phi) \\
W(\phi)\!&=&\!\phi\sqrt{|1-\phi^{ 2}|}\!\left(\dfrac{5}{8}-\dfrac{1}{4}\phi^2\right)\!+\!\frac{3}{8}\arcsin(\phi) \label{sp2}
\end{eqnarray}
\ees
for $n=1$ and $n=2$, respectively.

In Ref. \cite{first2}, we investigated the case $n = 2$. Here we use the Eq.~\eqref{eq8} and we obtain the general potential
\begin{eqnarray}
U(\phi)&=&\frac{2n-1}{2n}\left|1-\phi^2\right|^{n}\label{pott2}.
\end{eqnarray}
 
The choice \eqref{numvi} leads us to $\phi^{\prime}=\sqrt{|1-\phi^2|}$, which does not depend on $n$. It supports the compact solution
\begin{eqnarray}
\phi(x) &=& \left\{
\begin{array}{cl}
1 &  \mbox{ for } x<-\frac{\pi}{2}\\ 
\sin(x) & \mbox{ for } -\frac{\pi}{2}\leq x\leq \frac{\pi}{2} \\ 
-1 & \mbox{ for } x>\frac{\pi}{2}
\end{array} \right.
\end{eqnarray}
with the respective energy density
\begin{eqnarray}
\rho(x)=
 \left\{
\begin{array}{cll}
0 &,  \mbox{ for }& x<-\frac{\pi}{2}\\ 
\cos^{2n}(x) &, \mbox{ for }& -\frac{\pi}{2}\leq x\leq \frac{\pi}{2} \\ 
0 &, \mbox{ for }& x>\frac{\pi}{2}
\end{array} \right.
\end{eqnarray}
Here we note that the solution and energy density have a compact structure for all $n$. These kind of structures have also been studied in Refs \cite{ASW1,Bazeia:2010vb}.  The energy of the solution is 
\be
E=\frac{\sqrt{\pi}\,\Gamma{\left(n+\frac12\right)}}{\Gamma\left(n+1\right)}
\ee
For $n=1$ and $n=2$, the energy is $E=\pi/2$ and $E=3\pi/4$, respectively.

%%%%%%%%%%%%%%%%%%%%%%%%%%%%
\subsection{Linear Stability}
\label{sec1}

Let us now investigate linear stability.  We introduce a small fluctuation $\eta(x,t)$ about the static solution $\phi(x)$, that is, we write
\begin{equation}\label{eq10}
\phi(x,t)=\phi(x)+\eta(x,t)
\end{equation}
where $\phi(x)$ is solution of the static equation $(\ref{eq3})$. With this, we obtain, at first order in $\eta$,
\begin{eqnarray*}
\partial_\mu\left[\phi^{\prime 2n-2}\! \left(\partial^\mu\eta\!-\!2(n-1)\partial^\mu\phi\frac{\partial_\nu\phi\partial^\nu\eta}{\phi^{\prime 2}}\right)\right]\!+\!U_{\phi\phi}\eta=0.
\end{eqnarray*}	
Since $\phi=\phi(x)$ is static solution, we can assume that $\eta(x,t)=\eta_s(x)\cos(\omega t)$. Thus, we have
\begin{eqnarray}\label{eq11}
-(2n-1)[\phi^{\prime 2n-2}\eta_s^{\prime}]^{\prime}+U_{\phi\phi}|_s\eta_s=\omega^2\phi^{\prime 2n-2}\eta_s
\end{eqnarray}		
We follow Ref.~\cite{first1}, to rewrite the above equation as a Schroedinger-like equation. To do this, we introduce 
\begin{eqnarray}\label{eq12}
u(z)= (2n-1)^{\frac14} \phi^{\prime (n-1)}\,\eta_s\left(\sqrt{2n-1}\,\, z\right),
\end{eqnarray}
which allows writing $(\ref{eq11})$ as
\begin{equation}
-u_{zz}(z) + v(z) u(z)=\omega^2 u(z)
\end{equation}	
where
\begin{equation}\label{eq14}
v(z)=\frac{nU_{\phi\phi}}{\phi_z^{2n-2}}-\frac{n(n-1)}{(2n-1)}\frac{U_{\phi}^2}{\phi_z^{4n-2}}.
\end{equation}
Then, using Eqs.~$(\ref{eq7})$ and $(\ref{eq8})$ we obtain
\begin{equation}\label{spot}
v(z)=nW_\phi^{-\frac{2n-3}{2n-1}}W_{\phi\phi\phi}-\frac{n(n-2)}{2n-1}W_\phi^{-\frac{4(n-1)}{2n-1}}W_{\phi\phi}^2
\end{equation}
For kinklike solutions given by Eq.\eqref{spo1} the potential $v(z)$ is 
\be
v(z)=4(2n-1)n^2 - 2n(4n^2-1)\sech^2(\sqrt{2n-1}\,z) 
\ee

This is the modified P\"oschl-Teller potential \cite{PosT}. This potential supports the zero mode and other $2n-1$ 
bound states, with energies $E_k = (2n-1)k(4n - k)$, for $k =0,1,\ldots, 2n-1$. All the others
states of the model, with $w\geq 4n^2$ are not bounded.

It is interesting to see that we could assoociate the parameter $n$ with the number of bound states of the Schroedingre-like potential. As one knows, the number of bound states could in principe affect the rate of loss of the energy by radiation in dynamical processes, as for example, in a kink-antikink collision in comparation with the usual $\phi^4$ model (which is obtained with $n=1$). See, e.g., Ref.~ \cite{twob}.

 For the compacton solutions given by Eq.\eqref{numvi}, we get  
\begin{equation}
v(z)=\!n\lambda^2\!\left\{
\begin{array}{cl}
\infty &  \!\mbox{ for } z<-\frac{\pi}{2\lambda}\\ 
\!\!\!-n\!+\!(n\!-\!1)\sec^2\left(\lambda z\right)\! & \!\mbox{ for }\! -\frac{\pi}{2\lambda}\leq \! z \!\leq \frac{\pi}{2\lambda} \\ 
\infty & \!\mbox{ for } z>\frac{\pi}{2\lambda}
\end{array} \right.
\end{equation}
where $\lambda=\sqrt{2n-1}$. This is the P\"oschl-Teller potential \cite{PosT}. The interesting feature of this potential is that it only supports bound states, and for $k = 0, 1, 2, ...,$ the corresponding eigenvalues are given by $E_k = 4(2n-1)k(n+k)$. The radiation of energy for a collision between compactons is then completely different from the case of kinks.

%%%%%%%%%%%%%%%%%%%%%%%%%%%%%%			
\section{Twinlike Models}\label{other}

Let us now introduce a new family of twinlike models. We first recall that twinlike models are distinct models having the same solution, and the very same energy density. 
The main objective here is then to introduce a family of models which is twin to the family of models investigated in the previous section.
We follow the lines of Ref.~\cite{t3} and consider the theory
\begin{equation}\label{eq15}
{\cal L}=-U(\phi)F(Y),
\end{equation}
where $Y$ is defined as
\begin{equation}\label{eq16}
Y=-\frac{2^{n-1}}{n}\frac{X|X|^{n-1}}{U(\phi)}
\end{equation}
We note that if $n=1$,  we get back to the theory defined in \cite{t3}; also, for $F(Y)=1+Y$ we obtain the model introduced in Eq.~$(\ref{eq1})$ above.

This new model has the following equation of motion
\begin{equation}\label{eq17}
n\partial_\mu\left(U\frac{Y}{X}F_Y\partial^\mu\phi\right)-U_\phi F+YF_YU_\phi=0,
\end{equation}
and the energy-momentum tensor is given by
\begin{equation}
T_{\mu\nu}=g_{\mu\nu}U(\phi)F(Y)-nU\frac{Y}{X}F_Y\partial_\mu \phi\partial_\nu\phi.
\end{equation}
where $F_Y=dF/dY$.

As before, here we are interested in static field configurations; so, the equation of motion becomes
\begin{equation}\label{eq18}
-2n\left(U\frac{Y}{\phi^{\prime}}F_Y\right)^{\prime} +U_\phi F-YF_YU_\phi=0.
\end{equation}

We suppose that ${v_i \;( i = 1, 2, ..., n)}$ is a set of static
and uniform solutions of the equation of motion, meaning that $U^\prime(v_i)$ has to vanish. Also, we use the energy
density and take $U(v_i) = 0$ to make the energy itself vanish, for the static
and uniform solutions. Recall that the same conditions work for the standard model.

Since we are considering a new family of models, we
guide ourselves with the null energy condition (NEC),
that is, we take $T_{\mu\nu} n^\mu n^\nu \geq0$, where $n_\mu$ is a null vector, obeying $g_{\mu\nu} n^\mu n^\nu=0$. This restriction leads to
$F_Y\geq 0$, for the general field configuration $\phi(x,t)$ which is supposed to solve the equation of motion \eqref{eq17}.
Moreover, for static solutions, the energy-momentum tensor gives
\bes
\begin{eqnarray}
T_{00}&=&UF, \\
T_{11}&=&-UF+2nYF_YU.
\end{eqnarray}
\ees
The Eq.~ $(\ref{eq18})$ can be integrated once to give
\begin{equation}\label{eq20}
2nYF_Y-F=\frac{C}{U}.
\end{equation}
Again, $C$ is a constant identified with the stress tensor $T_{11}$. Furthermore, we have
\begin{equation}\label{eq21}
Y=\frac{1}{2n}\frac{\phi^{\prime 2n}}{U(\phi)}
\end{equation} 
The Eq.~\eqref{eq20} can be written in the form
\begin{equation}\label{eq22}
\phi^{\prime 2n}=2nG\left(\frac{C}{U}\right) U(\phi)
\end{equation}
where $G$ is a function with inverse $G^{-1}(Y)=2nYF_Y-F$. 

For stressless solutions, that is, for $C=0$, we have that $2nYF_Y\!=\!F$ and if we assume that $G(0)=c$, with $c$ constant, real, we find that $Y=c$. With this result, we can rewrite Eq.~$(\ref{eq22})$ in the form
\begin{equation}\label{eq23}
\phi^{\prime 2n}=2nc U(\phi)
\end{equation}

Here we note that the solution $\phi(x)$ of this equation is the same solution $\phi_s(x)$ of the Eq.~\eqref{eq4}, which appears for the previous model, with the position changed as $x \to \sqrt{m} x$, with $m=[c(2n-1)]^{1/n}$. This means that we can write
\be
\phi(x) = \phi_s(\sqrt{m}\,x), \label{ppp12}
\ee
and now the thickness of the solution is given by
\be
\delta=\delta_s/\sqrt{m} .
\ee
Thus, the solution is thicker or thinner, depending on the
value of $c$ being lesser or greater than unit. We also note that
$c$ cannot be negative;  and more, only stressless solutions have the specific form, given by Eq. \eqref{ppp12}.

The energy density of the stressless solution \eqref{eq23} gets to the form
\begin{equation}\label{eq24}
\rho(x)=F(c)U(\phi(x))
\end{equation}
The energy is then $E=F(c)\int_{-\infty}^\infty  dx\, U(\phi(x))$, or better,
\ben
E&=&\frac{F(c)}{[c(2n-1)]^{1/(2n)}} \int_{-\infty}^\infty  dy\, U(\phi_s(y))\\&=&\frac{(2n-1)F(c)}{2n[c(2n-1)]^{1/(2n)}}E_s
\een
where $E_s$ is the energy for $F=1+Y$. 

The solutions with a non-vanishing $T_{11}$ are
different from the corresponding solutions of the previous model, because they do not have the form given by Eq.~\eqref{ppp12}.
We recall that for $T_{11} = C$, only the stressless solutions are stable \cite{first1}. Usually, the energy of the other
possible solutions are divergent, and the solutions have oscillatory
or divergent profiles. We find the same behavior in the standard model.

Now we use again the formalism introduced in \cite{first1,first2} to rewrite the energy density for the generalized model. Assuming that the Eq.~$(\ref{eq6})$ is valid, we have that
\begin{equation}\label{eq25}
\phi^{\prime}= F_Y^{-\frac{1}{2n-1}} W_\phi^{\frac{1}{2n-1}} 
\ee
The potential is given by
\be\label{eq26a}
U(\phi)=\frac{F_Y^{-\frac{2n}{2n-1}}}{2n Y }W_\phi^{\frac{2n}{2n-1}}
\end{equation}
and we obtain the energy density in the form
\begin{equation}\label{eq27}
\rho(x)=\frac{F\,F_Y^{-\frac{2n}{2n-1}}}{2n Y } W_\phi^{\frac{2n}{2n-1}}.
\end{equation}
Now, using $(\ref{eq20})$ with $C=0$ we have
\begin{equation}\label{eq28}
\rho(x)=F_Y ^{-\frac{1}{2n-1}} W_\phi^{\frac{2n}{2n-1}}.
\end{equation}
For $m=1$, i. e., 
\be \label{cccc}
c=\frac{1}{2n-1},
\ee
we have to impose 
\bes\label{twinconditions}
\be \label{LLL12}
F_Y((2n-1)^{-1})=1
\ee
in order to make the Eqs.~$\eqref{eq25}$ and $(\ref{eq28})$ identical to the Eqs.~$\eqref{eq7}$ and $(\ref{eq9})$, respectively. This also imposes that 
 \be \label{eqrrr21}
 F((2n-1)^{-1})=\frac{2n}{2n-1}
 \ee
\ees
The Eqs.~\eqref{LLL12} and \eqref{eqrrr21} are the general restrictions on $F(Y)$, to make the model twin of the previous model.

%%%%%%%%%%%%%%%%%%%%%
\subsection{Linear Stability}\label{sec2}
	
Let us again investigate linear stability by introducing small fluctuations $\eta(x,t)$ in the static solution $\phi(x)$, as done in Sec.~\ref{sec1}. Using $(\ref{eq10})$ in $(\ref{eq17})$ we obtain

\begin{eqnarray}
-\partial_\mu\!\Bigg[\phi^{\prime 2n-2}\!\Bigg(-\!2(n\!-\!1)F_Y\partial^\mu\phi\frac{\partial_\alpha\phi \partial^\alpha\eta}{\phi^{\prime 2}}\nonumber \\+F_Y\partial^\mu\eta+F_{YY}\Delta Y\partial^\mu\phi\Bigg)\!\Bigg]=
U_{\phi\phi}(F-YF_Y)\eta\nonumber \\+U_\phi(F-YF_Y)_Y\Delta Y,
\end{eqnarray}

where
\begin{equation}
\Delta Y \equiv Y\left(-\frac{2n}{\phi^{\prime 2}}\partial_\beta\phi\partial^\beta\eta -\frac{U_\phi}{U}\eta\right).
\end{equation}

Taking $\eta(t,x)=\eta_s(x)\cos(\omega t)$ we get 
\begin{eqnarray}\label{eq30}
-\left[q(x)\left[2nF_{YY}Y +(2n-1)F_Y\right]\eta_s^{\prime} \right]^{\prime}=\nonumber\\
\left[U_{\phi\phi}(YF_Y-F)-\left(\phi^{\prime 2n-1}F_{YY}Y\frac{U_\phi}{U}\right)^{\prime}-\right.\nonumber\\
\left.-F_{YY}Y^2\frac{U_\phi^2}{U}\right]\eta_s+\omega^2 F_Y q(x) \eta_s ,
\end{eqnarray}
where
$q(x)\equiv \phi^{\prime 2n-2}$.

In the case of a stressless solution we can use the Eq.~(\ref{eq20}) with $C=0$ to transform $(\ref{eq30})$ in the form
\begin{equation}\label{eq31}
-\left[q(x)\eta_s^{\prime} \right]^{\prime}+U_{\phi\phi}Y\eta_s=\frac{\omega^2}{A^2}q(x) \eta_s.
\end{equation}
where
\begin{equation}\label{eq32}
A^2=\frac{2nF_{YY}Y +(2n-1)F_Y}{F_Y}.
\end{equation}
We note that $A$ is constant for a stressless solution. Also, it is required that $A$ is positive, if we want to ensure hyperbolicity of the differential equation.

Again, we introduce the suggested exchange of variables, 
\ben
u(z)=F_Y^{\frac12} A^{\frac12} \phi^{\prime n-1}\eta_s(A z) 
\een
Here we get
\begin{equation}
-u_{zz}(z) + v_2(z) u(z)=\omega^2 u(z),
\end{equation}	
where 
\begin{equation}\label{eq33}
v_2(z)=nA^2Y\left(\frac{U_{\phi\phi}}{\phi_z^{2n-2}}-(n-1)Y\frac{U_{\phi}^2}{\phi_z^{4n-2}}\right)_{\phi=\phi_s(A z)}
\end{equation}
Now, using $(\ref{eq25})$ and $(\ref{eq26a})$ we obtain
\begin{equation}
v_2(z)\!=\!\frac{A^2 F_Y^{-\frac{2}{2n\!-\!1}}}{2n\!-\!1}\!\left(\!n\frac{W_{\phi\phi\phi}}{W_\phi^{\frac{2n-3}{2n-1}}}\!-\!\frac{n(n\!-\!2)}{2n\!-\!1} \frac{W_{\phi\phi}^2}{W_\phi^{\frac{4(n\!-\!1)}{2n\!-\!1}}}\!\right)
\end{equation}
We note that if we impose the twin conditions \eqref{twinconditions}, we obtain $A^2=2n-1+2n(2n-1)^{-1} F_{YY}$. With this, we obtain the relation $v_2(x)=v(x)$, if we choose 
\be \label{eqrrr2}
 F_{YY}((2n-1)^{-1})=0
 \ee

Thus, we see that it is possible to find twinlike models  starting from non standard theories. This is a new result, since before one usually started from a standard field theory, in order to construct the related twinlike model.

%%%%%%%%%%%%%%%%%%%%%%%%%%%%%%
\subsection{Examples}

Let us now specify the function $F(Y)$ in order to illustrate how the formalism introduced above works. The first model  we consider is 
\begin{equation}\label{FYab}
F(Y) = a + bY|Y|^{k-1}
\end{equation}
Here we suppose that $k \geq 1$ and $a,b$ are real number. 

To consider the stressless solution, we write
\begin{equation}
c=\left(\frac{a}{2n kb -b}\right)^{1/k}
\end{equation}
We note that 
\be F_Y(c)=bk\left(\frac{a}{2nkb-b}\right)^{\frac{k-1}{k}} {\rm and} \,\,
\,  F(c)=\frac{2nka}{2nk-1} \ee
The conditions \eqref{twinconditions} give
\be
a=\frac{2nk-1}{k(2n-1)}   \mbox{\;\;\; and \;\;\;} b=\frac{(2n-1)^{k-1}}{k}
\ee
The function given by Eq. \eqref{FYab} can be written as
\be
F(Y)=\frac{2nk-1}{k(2n-1)} +\frac{(2n-1)^{k-1}}{k}Y|Y|^{k-1}
\ee
and so we get to the Lagrange density
\begin{equation}
{\cal L}= (2n-1)^{k-1}\frac{2^{(n-1)k}}{kn^{k}}\frac{X|X|^{nk-1}}{U^{k-1}}-\frac{2nk-1}{k(2n-1)}U(\phi)
\end{equation}
To investigate linear stability we have to consider
\begin{equation}
A^2=2n k-1
\end{equation}
and so we note that the condition \eqref{eqrrr2} requires that $k=1$. However, in this case the twin theory is identically to the original model.

We now introduce the second family of models, which obeys the strong condition. Let us consider the following function 
\begin{equation}
F(Y) = 1+Y+G\left(Y-\frac{1}{2n-1}\right)
\end{equation}
For the three conditions (same solution, same energy density and same stability behavior) to be valid, one imposes that
\be
G(0)=G^\prime(0)=G^{\prime\prime}(0)=0.
\ee
We can write a general function which obeys these three conditions; it has the form
\begin{equation}
F(Y) = 1+Y+\sum_{i>2} \beta_i  \left(Y-\frac{1}{2n-1}\right)^{i}
\end{equation}
where  all the $\beta_i$ are real parameters.

%%%%%%%%%%%%%%%%%%%%%
\section{Braneworld models}\label{brane}

Let us now investigate how the twinlike models studied above can be used to represent generalized braneworld models. Here we follow the lines of Ref.~\cite{fb}. In this context, we consider an action in five dimensions that describe gravity coupled to a scalar field in the form
\begin{equation}
S=\int d^5x\sqrt{g}\left(-\frac{1}{4}R+{\cal L}(\phi,X)\right)
\end{equation}
where we are using $4\pi G=1$ and also
\begin{equation}
X=\frac{1}{2}\nabla_M \phi \nabla^M \phi 
\end{equation}
with $M,N=0,1,2,3,4$ running on the five-dimensional spacetime. The equation of motion which we obtain is given by
\begin{equation}
G^{NM}\nabla_N\nabla_M\phi+2X{\cal L}_{X\phi}-{\cal L}_{\phi}=0,
\end{equation}
where
\begin{equation}
G^{NM}={\cal L}_{XX}\nabla^M\phi\nabla^N\phi+g^{MN}{\cal L}_{X}.
\end{equation}
The energy-momentum tensor $T_{MN}$ has the form
\begin{equation}T_{MN}=-g_{MN}{\cal L}+{\cal L}_X\nabla_M\phi\nabla_N\phi
\end{equation}
The line element of the five-dimensional spacetime can be written as $ds^2= e^{2{\cal A}}\eta_{\mu\nu}dx^\mu dx^\nu-dy^2$, where ${\cal A}$ is used to describe the warp factor. We suppose that both ${\cal A}$ and $\phi$ are static, such that they only depend on the extra dimension $y$, that is, ${\cal A} ={\cal A}(y)$ and $\phi = \phi(y)$. In this case, the equation of motion for the scalar field reduces to
\begin{equation}\label{eq34}
(2X{\cal L}_{XX}+{\cal L}_{X})\phi^{\prime \prime}-(2X{\cal L}_{X\phi}-{\cal L}_{\phi})=-4{\cal L}_{X}\phi^{\prime} {\cal A}^{\prime}
\end{equation}
Moreover, from the Einstein equations we get
\bes
\begin{eqnarray}
{\cal A}^{\prime\prime}&=&\frac{4}{3}X{\cal L}_X\label{eq35}\\
{\cal A}^{\prime 2}&=&\frac{1}{3}({\cal L}-2X{\cal L}_X)\label{eq36}
\end{eqnarray}
\ees
where $X=-\phi^{\prime 2}/2$ for static configuration, as before. 

To get to the first-order framework, we suppose that
\begin{equation}\label{eq37}
{\cal A}^{\prime} =-\frac{1}{3}W(\phi)
\end{equation}
In this case, the Eqs.~(\ref{eq35}) and (\ref{eq36}) lead us to, respectively,
\begin{eqnarray}
\phi^{\prime}{\cal L}_X&=&\frac{1}{2}W_\phi, \label{eq38}\\
{\cal L}-2X{\cal L}_X&=&\frac{1}{3}W^2(\phi).\label{eq39}
\end{eqnarray}

In the case of theory $(\ref{eq1})$ the equation of motion $(\ref{eq34})$ becomes
\begin{equation}\label{eq40}
(2n-1)\phi^{\prime 2n-2}\phi^{\prime \prime}+4\phi^{\prime 2n-1} {\cal A}^{\prime}=U_\phi.
\end{equation}
We use Eq.~(\ref{eq38}) to write
\begin{equation}\label{eq41}
\phi^{\prime} =2^{\frac{1}{1-2n}} W_\phi^{\frac{1}{2n-1}},
\end{equation}
and the potential
\begin{equation}\label{eq42}
U(\phi)=\frac{2n-1}{n 2^{\frac{4n-1}{2n-1}}}W_\phi^{\frac{2n}{2n-1}}- \frac{1}{3}W^2(\phi),
\end{equation}
and the energy density
\begin{equation}\label{eq43}
T_{00}=e^{2{\cal A}}\left({2^{\frac{2n}{1-2n}}}W_\phi^{\frac{2n}{2n-1}}- \frac{1}{3}W^2(\phi)\right)
\end{equation}
Now we have to find the twin model. For this, let us consider a scalar field theory governed by the following Lagrange density
\begin{equation}\label{eq44}
{\cal L}=-U(\phi)F(Y)+f(\phi)
\end{equation}
where $Y$ was defined in $(\ref{eq16})$ and $f(\phi)$ is to be determined. We use the Eqs.~$(\ref{eq38})$ and $(\ref{eq39})$ to write, respectively
\begin{equation}\label{eq45}
\phi^{\prime}=\frac{1}{(2F_Y)^{\frac{1}{2n-1}}}W_\phi^{\frac{1}{2n-1}},
\end{equation}
and
\begin{equation}\label{eq46}
\phi^{\prime 2n}=\frac{F}{F_Y}U(\phi),
\end{equation}
where we have used $f(\phi)=W^2/3$.

We can so to rewrite \eqref{eq46} in the form
\begin{equation}
F=2nYF_Y
\end{equation}

The Lagrange density of the twin brane model then has the following form
\begin{equation}\label{eq47}
{\cal L}=-U(\phi) F(Y)+\frac{1}{3}W^2(\phi)
\end{equation}
Moreover, the energy density is
\begin{equation}\label{eq48}
T_{00}=e^{2{\cal A}}\left[\frac{2^{\frac{2n}{1-2n}}}{F_Y^{\frac{1}{2n-1}}}W_\phi^{\frac{2n}{2n-1}}-\frac{1}{3}W^2(\phi)\right]
\end{equation}
which exactly reproduces the previous expression (\ref{eq43}) if 
\bes\label{twiiii}
\be
F_Y(c)=1\ee and, consequently, \be F(c)=2n(2n-1)^{-1}.\ee\ees Thus, the two models have the same solution,
with the very same energy density. And these are the two conditions required for the models to be twinlike models. 

\subsection{Brane Stability}

The investigation of the linear stability of the braneworld model can be done following Ref. \cite{fb}. The metric is perturbed in the form
\be
ds^2=e^{2A(y)}\left(\eta_{\mu\nu}+h_{\mu\nu}(y,x)\right) dx^\mu dx^\nu -dy^2
\ee
and the scalar field in the form
\be
\phi=\phi(y)+\tilde \phi(y,x)
\ee
For the starting model, given by Eq. \eqref{eq1}, the first order contributions to the energy-momentum tensor are
\begin{eqnarray}
\overline{T}_{\mu\nu}^{(1)}\!&=&\!\frac{\eta_{\mu\nu}e^{2 {\cal A}}}{3}W_\phi\Big[(n-1)\tilde{\phi}^{\prime}\!-\!{2^{\frac{1}{1-2n}}}W_\phi^{\frac{2-2n}{2n-1}}W_{\phi\phi}\tilde{\phi}+\nonumber\\
\nonumber
\!\!&+&\!\!\frac{4}{3}W\tilde{\phi}\Big]\!-\!2e^{2 {\cal A}}h_{\mu\nu}\!\left[{2^{\frac{1-4n}{2n-1}}}W_\phi^{\frac{2n}{2n-1}}-\frac{1}{3}W^2\right]\nonumber\\
\overline{T}_{\mu 4}^{(1)}&=&\frac{1}{2}W_\phi\nabla_\mu\tilde{\phi}\\
\overline{T}_{44}^{(1)}&=&\frac{2^{\frac{1}{1-2n}}}{3}{}W_\phi^{\frac{1}{2n-1}}W_{\phi\phi}\tilde{\phi}-\frac{4}{9}W_\phi W\tilde{\phi}\nonumber +\frac{2n-1}{3}W_\phi\tilde{\phi}^{\prime}
\end{eqnarray}
The first order contributions to the Einstein equations are
\begin{eqnarray}\label{eqeqeqq}
&&\!\!\!e^{2{\cal A}}\left(\frac{1}{2}\partial_y^2-\frac{2}{3}W\partial_y\right)h_{\mu\nu}-\frac{1}{6}\eta_{\mu\nu}e^{2{\cal A}}W\partial_y(\eta^{\alpha \beta}h_{\alpha\beta})+\nonumber\\&&
-\frac{1}{2}\eta^{\alpha\beta} (\partial_\mu\partial_\nu h_{\alpha\beta} - \partial_\mu\partial_\alpha h_{\nu\beta} -\partial_\nu\partial_\alpha h_{\mu\beta})\nonumber\\
&&\!\!\!=\frac{4e^{2{\cal A}}\eta_{\mu\nu}}{3}W_\phi\!\left[\frac{(n-1)}{2}\tilde{\phi}^{\prime}\!-\!\frac{W_\phi^{\frac{2-2n}{2n-1}}W_{\phi\phi}}{2^{\frac{1}{2n-1}}}\tilde{\phi}\!+\!\frac{4W}{3} \tilde\phi\right]
\end{eqnarray} 
and 
\begin{eqnarray}
\frac{1}{2}\eta^{\alpha\beta} \partial_y(\partial_\alpha h_{\mu \beta}-\partial_\mu h_{\alpha\beta})
&=&\frac{1}{2}W_\phi\partial_\mu\tilde{\phi}\\
-\frac{1}{2}\left(\partial^2_y+\frac{2}{9}W^2\partial_y\right)\eta^{\alpha\beta}h_{\alpha\beta}\nonumber
&=&\frac{1}{3}\frac{1}{2^{\frac{1}{2n-1}}}W_\phi^{\frac{1}{2n-1}}W_{\phi\phi}\tilde{\phi}\\-\frac{4}{9}W_\phi W\tilde{\phi}\!\!\!\!\!\!\!\!\!\!\!\!&&+\frac{(2n+1)}{3}W_\phi\tilde{\phi}^{\prime}
\end{eqnarray}

The equation of motion for the scalar field leads to
\begin{eqnarray}
&&W_\phi^{\frac{2n-2}{2n-1}}e^{2{\cal A}}\square \tilde{\phi}-(2n-1)\left[W_\phi^{\frac{2n-2}{2n-1}}\tilde{\phi}^{\prime}\right]^{\prime}+\nonumber\\
&&+\frac{4(2n-1)}{3}WW_\phi^{\frac{2n-2}{2n-1}}\tilde{\phi}^{\prime}+
\frac{2^{\frac{2}{1-2n}}}{2n-1}W_\phi^{\frac{2-2n}{2n-1}}W_{\phi\phi}^2\tilde{\phi}\nonumber\\
&&+\frac{1}{2^{\frac{2}{2n-1}}}W_\phi^{\frac{1}{2n-1}}W_{\phi\phi\phi}\tilde{\phi}-\frac{2^{\frac{4n-3}{2n-1}}}{3}W_{\phi\phi} W\tilde{\phi}\nonumber\\
&&-\frac{2^{\frac{4n-3}{2n-1}}}{3}W_\phi^2\tilde{\phi}
=\frac{1}{2^\frac{1}{2n-1}}W_\phi\eta^{\alpha\beta}h_{\alpha\beta}
\end{eqnarray}

For the general model \eqref{eq47}, after substituting the two twin conditions \eqref{twiiii}, one is led to following set of equations: i) the energy-momentum components:
\begin{subequations}
\begin{eqnarray}
\overline{T}_{\mu\nu}^{(1)}&=&\frac{\eta_{\mu\nu}e^{2 {\cal A}}}{3}W_\phi\Bigg[\left(\frac{nF_{YY}}{2n-1}+n-1\right)\tilde{\phi}^{\prime}\nonumber\\
&-&\!\!\!\left(1+\frac{nF_{YY}}{(2n-1)^2}\right) 2^\frac{1}{1-2n}
W_\phi^{\frac{2-2n}{2n-1}}W_{\phi\phi} \tilde{\phi}+\frac{4}{3} W\tilde{\phi}\Bigg]-\nonumber\\
&-&2e^{2 {\cal A}}h_{\mu\nu}\left[2^{\frac{1-4n}{2n-1}}W_\phi^{\frac{2n}{2n-1}}-\frac{1}{3}W^2\right]\\
\overline{T}_{\mu 4}^{(1)}&=&\frac{1}{2}W_\phi\nabla_\mu\tilde{\phi}\\
\overline{T}_{44}^{(1)}&=&\frac{2^\frac{1}{1-2n}}{3}\left(1-2n\frac{F_{YY}}{(2n-1)^2}\right)W_\phi^{\frac{1}{2n-1}}W_{\phi\phi}\tilde{\phi} \nonumber \\
&-&\frac{4}{9}W_\phi W\tilde{\phi}+\left(\frac{2n}{3}\frac{F_{YY}}{2n-1}+\frac{2n-1}{3}\right)W_\phi\tilde{\phi}^{\prime}
\end{eqnarray}
\end{subequations}
ii) the Einstein equations:
\begin{eqnarray}
&&e^{2{\cal A}}\left(\frac{1}{2}\partial_y^2-\frac{2}{3}W\partial_y\right)h_{\mu\nu}-\frac{1}{6}\eta_{\mu\nu}e^{2{\cal A}}W\partial_y(\eta^{\alpha \beta}h_{\alpha\beta})+\nonumber\\
&&-\frac{1}{2}\eta^{\alpha\beta} ( \partial_\mu\partial_\nu h_{\alpha\beta} - \partial_\mu\partial_\alpha h_{\nu\beta} -\partial_\nu\partial_\alpha h_{\mu\beta})\nonumber\\ &&
=\frac{4e^{2{\cal A}}\eta_{\mu\nu}}{3}W_\phi\Bigg[\frac{1}{2}\left(n\frac{F_{YY}}{2n-1}+n-1\right) \tilde{\phi}^{\prime}\\&&-\frac{1}{2^{\frac{2n}{2n-1}}}\left(2+\frac{nF_{YY}}{(2n-1)^2}\right)W_\phi^{\frac{2-2n}{2n-1}}W_{\phi\phi} \tilde{\phi}+\frac{4}{3}W\tilde{\phi}\Bigg]\nonumber
\end{eqnarray} 
and 
\begin{subequations}
\begin{eqnarray}
\frac{1}{2}\eta^{\alpha\beta} \partial_y(\partial_\alpha h_{\mu \beta}-\partial_\mu h_{\alpha\beta})&=&\frac{1}{2}W_\phi\partial_\mu\tilde{\phi}\\-\frac{1}{2}\left(\partial^2_y+\frac{2}{9}W^2\partial_y\right)\eta^{\alpha\beta}h_{\alpha\beta} &=&-\frac{4}{9}WW_\phi\tilde{\phi}+\nonumber\\
+ \frac{1}{3} \left(\frac{1-2n \frac{F_{YY}}{(2n-1)^2}}{2^{\frac{1}{2n-1}}}\right)\!\!\!\!\!\!&&\!\!\!\!\!\!W_\phi^{\frac{1}{2n-1}}W_{\phi\phi}\tilde{\phi} \nonumber
\\+\frac{1}{3}\left(\frac{2nF_{YY}}{2n-1}+2n+1\right)\!\!\!\!\!\!&&\!\!\!\!\!\!W_\phi\tilde{\phi}^{\prime}
\end{eqnarray}
\end{subequations}
iii) and the scalar field equation:
\begin{eqnarray}
&&W_\phi^{\frac{2n-2}{2n-1}}e^{2{\cal A}}\square \tilde{\phi}-\left(2n\frac{F_{YY}}{2n-1}+2n-1\right)\left[W_\phi^{\frac{2n-2}{2n-1}}\tilde{\phi}^{\prime}\right]^{\prime}\nonumber\\
&&+\frac{4}{3}\left(2n\frac{F_{YY}}{2n-1}+2n-1\right) W_\phi^{\frac{2n-2}{2n-1}}W\tilde{\phi}^{\prime}+\nonumber\\
&&+
\left(\frac{1+2n\frac{F_{YY}}{(2n-1)^2}}{2^{\frac{2}{2n-1}}(2n-1)}\right)W_\phi^{\frac{2-2n}{2n-1}}W_{\phi\phi}^2\tilde{\phi}+ \nonumber
\\&&
+\left(\frac{2n\frac{F_{YY}}{(2n-1)^2}+1}{2^{\frac{2}{2n-1}}}\right)W_\phi^{\frac{1}{2n-1}}W_{\phi\phi\phi}\,\tilde{\phi}-\nonumber\\
&&-\frac{2^{\frac{4n-3}{2n-1}}}{3}\left(\frac{2nF_{YY}}{(2n-1)^2}+1\right)W_{\phi\phi}W\tilde{\phi}-\frac{2^{\frac{4n-3}{2n-1}}}{3}W_\phi^2\tilde{\phi}\nonumber
\\&&=\frac{1}{2^{\frac{1}{2n-1}}}W_\phi\eta^{\alpha\beta}h_{\alpha\beta}
\end{eqnarray}

We see that only for 
\be\label{strongcond} F_{YY}=0,
\ee 
the set of equations is equivalent to that corresponding to the starting model. As we know, the study of stability is not a trivial task \cite{DeWolfe:2000xi}; however, we can assure here that, using the three conditions -- same solution, same energy density, and the strong condition \eqref{strongcond} -- the linear stability of the two models are the same. 

In the gravity sector, we can simplify the investigation of stability considering the transverse traceless components of metric fluctuations
\ben
\bar{h}_{\mu\nu}=\left(\frac{1}{2}(\pi_{\mu\alpha}\pi_{\nu\beta}+\pi_{\mu\beta}\pi_{\mu\alpha})-\frac13 \pi_{\mu\nu}\pi_{\alpha\beta}\right)h^{\alpha\beta}
\een
where $\pi_{\mu\nu}=\eta_{\mu\nu} -\partial_\mu \partial_\nu/\Box.$ Indeed, we can check that Eq.~\eqref{eqeqeqq} reduces to the known equation
\ben
\left(\partial^2_y + 4 {\cal A}^\prime \partial_y - e^{-2{\cal A}}\Box\right)\bar{h}_{\mu\nu}=0.
\een
The next steps are known: we introduce the $z$-coordinate in order to make the metric conformally flat, with $dz=e^{-A(y)}dy$ and we write
\ben
H_{\mu\nu}(z)=e^{-ipx}e^{3/2{\cal A}(z)}\bar{h}_{\mu\nu}
\een
In this case, the 4-dimensional components of ${\bar h}_{\mu\nu}$ obey the Klein-Gordon equation and the metric fluctuations of the brane solution lead to the Schr\"odinger-like equation
\ben
\left[-\partial_z^2 + U(z) \right]H_{\mu\nu} = p^2 H_{\mu\nu}
\een
where 
\ben
U(z)=\frac{9}{4}{\cal A}^{\prime2}(z) + \frac32 {\cal A}^{\prime\prime}(z).
\een
We note that the stability behavior in the gravity sector only depends on the warp factor ${\cal A}$, so the two first conditions to make the two models twins -- same solution and same energy density -- are necessary for the two models to have the same stability behavior in the gravity sector. 

Therefore, we can write the following two important conclusions concerning stability of the two general models, described by Eq.~\eqref{eq1} and by Eq.~\eqref{eq44}, in the braneworld context: i) stability in the gravity sector is controlled by the warp factor, so the two first conditions for the models to be twins -- explicitly, same solution and same energy density -- lead the two twin models with the very same stability behavior; ii) stability in the scalar field sector are in general different, but the strong condition -- as given by Eq.~\eqref{strongcond} -- makes the two models to have the very same stability behavior in the scalar field sector too.

The above results show that the modifications proposed in the current work are robust and may be of direct interest to high energy physics.

%%%%%%%%%%%%%%%%%%%%
\section{Conclusions}
\label{end} 
In this work we introduced another route to construct twinlike models, now starting from a generalized  model, and generating another generalized model. We did this investigating several examples, showing that the procedure is generic and work for a diversity of models. 

To make the investigation stronger, in this work we also discussed the case of branes with warped geometry, in the scenario with a single extra dimension of infinite extent. Here we also investigated how the two first conditions for the models to be twins, namely, the same solution and the same energy density, and the extra condition, which we called strong condition, enter the game when one investigates stability. The result is that stability in the gravity sector is controlled by the warp factor, so it requires that the models are twins, that is, that they present the same solution, with the same energy density. In the scalar field sector, however, stability has also the same behavior if we includes the third condition, the strong condition \eqref{strongcond}.

The procedure seems to be robust, working for several distinct models, valid both in the flat and curved spacetime, in the last case for a braneworld model with a single extra dimension of infinite extent.

\end{document}